# Ultracold Neutrons

K. Kirch[1,2,a] B. Lauss[1,b]
[1] Laboratory for Particle Physics, Paul Scherrer Institut, CH-5232 Villigen-PSI, Switzerland, [2] Institute for Particle Physics and Astrophysics, ETH Zürich, Zürich, Switzerland

March 19, 2020

## Abstract

Ultracold neutrons (UCN) are free neutrons that can be stored in experimental setups for several minutes. Some of the most important properties of the neutron, such as its tiny permanent electric dipole moment and its beta decay lifetime, are best measured with UCN. Also searches for well motivated but yet unknown, hypothetical additional interactions are being pursued with UCN. Such measurements in the field of low-energy, precision physics may have far reaching implications from particle physics to cosmology. Most experiments are statistics limited and need high-intensity UCN sources. The UCN source at PSI is at the forefront of the field and home to the international nEDM collaboration and its world-leading search for the neutron electric dipole moment. This article aims at giving an overview of the fascinating research using ultracold neutrons emphasizing on activities at PSI including various physics side-analyses which were pioneered by the nEDM collaboration.

PACS. XX.XX.XX No PACS code given

# 1 Ultracold neutrons

Fundamental neutron physics is new to the Swiss Neutron Science Society. Measurements with considerable impact on the field of fundamental interactions and symmetries, often in a context of nuclear and particle physics, can be performed with neutrons, and in particular with ultracold neutrons.

The neutron itself was predicted in the 1920s by Rutherford and discovered in 1932 by Chadwick. In terms of basic constituents of matter, in many sense, this completed and at the same time initiated a revolution of microscopic physics at the time, allowing to develop and implement modern theories of atomic and nuclear physics, and also starting particle physics. Within the next few years, in particular Fermi and collaborators, but also others, performed many basic experiments and found proper descriptions for a large body of neutron physics and concepts, including moderation and total reflection of thermal neutrons under grazing angles by the neutron optical potential ('Fermi pseudo-potential') of materials, see e.g. [1–3]. While the idea of neutron storage for sufficiently slow neutrons is often also attributed to Fermi, it has been published for the first time by Zel'dovich [4] in 1959. The discovery of ultracold neutrons (UCN) occurred in 1969, independently and almost at the same time in Dubna [5] and in Munich [6]. UCN have thus just celebrated their 50th anniversary last year. Two excellent textbooks on UCN physics are available and recommended to the reader [7,8].

UCN are free neutrons with very low kinetic energies and very long de Broglie wavelengths that can be stored in evacuated containers with material walls. While thermal neutrons can be reflected under grazing angles, UCN undergo total reflection at any angle of incidence on suitable materials with sufficiently large Fermi potential. Kinetic energies of UCN are typically of order 100 neV, their velocity a few m/s and their wavelengths in the range of 100 nm, not too far from visible light. In many respects, UCN behave like an ideal non-interacting gas. Stored for sufficiently long time in containers, they can reach mechanical equilibrium, filling the available phase space. However, it is important to realize that they are not in thermal equilibrium with their surrounding. In collisions with the containment, the UCN energy is conserved, i.e. the kinetic energies before and after wall collisions are the same. While the temperature one could calculate for UCN amounts to millikelvins, they can be stored in room-temperature containers.[1]

The possibility to confine UCN in traps and to use them for long times in experiments is ultimately limited only by the neutron beta decay lifetime of about 15 minutes.[2] It is amazing that the typical energies of different UCN interactions are on the same order of magnitude as their kinetic energy: the strong interaction potential of many wall materials is larger than 100 neV, the magnetic moment interaction amounts to 60 neV/T and the gravitational potential is 102neV/m. This coincidence allows for confinement options using evacuated material bottles, magnetic bottles and combined systems open to the top, using gravitational vertical confinement[3].

---

[1] This is like for cold neutron reflection under grazing angles from neutron guide surfaces. The repulsive interaction on the wall material is the coherent interaction with very many nuclei of the wall (compare wavelength to typical lattice spacing). While the individual nuclei of the wall material will be in motion around their equilibrium position, according to the temperature and the phonon spectrum of the material, they will be almost perfectly at rest in the laboratory when averaged over the many interacting nuclei and together of almost infinite mass compared to the neutron. Therefore, the neutron kinetic energy is not changed in a wall collision, the neutron gas not heated and nominally at mK temperature in room temperature containers.

To date UCN are mostly used for fundamental physics studies determining properties of the neutron itself (its lifetime, magnetic and electric dipole moments, electric charge neutrality) or of fundamental interactions (e.g. weak decay, gravitational interaction, search for additional e.g. spin-dependent or Lorentz violating interactions).

Recalling the beginning, the 'appearance' of the neutron helped to solve the crisis in the understanding of microscopic atomic and nuclear physics. Perhaps, again, the neutron could play a central role in the resolution of crises of our understanding. Fundamental physics today faces many open questions, in particular in the marriage of microscopic physics and cosmological observations. The cosmological standard theory starts with a big-bang generating equal amounts of matter an anti-matter. So far, we observe no primordial antimatter at all while the matter of interstellar gas and stars, and us, is definitely there. This calls for an asymmetry between matter and antimatter ('CP violation') [9] beyond the one known in the Standard Model of particle physics (SM). In addition, from astrophysical observations, there is overwhelming evidence for the existence of another form of matter that we cannot observe directly so far but only due to its gravitational interaction, so-called Dark Matter[4]. Our best theory of the microscopic world so far, the SM, is a quantum field theory and as such not compatible with classical gravity. The SM is relying safely on very fundamental and testable assumptions, such as Lorentz-Invariance (LI), but it offers not enough CP violation and no candidate particles that could constitute the Dark Matter.

UCN allow for various most sensitive experiments that can address aspects of these open issues of modern physics spanning the astounding range from the smallest microscopic to the largest cosmological scales. PSI is home to some of the leading efforts in this field as will be illustrated in this article.

---

[2] Besides the coherent interaction with the wall's nuclei, UCN can also scatter incoherently on individual nuclei. This can lead to spin-flips or up-scattering to higher energies which is usually equivalent to a loss of UCN as they can then leave the confinement. A tiny fraction might stay as 'heated UCN'. Also neutron absorption on nuclei is possible. It can be a significant experimental challenge to effectively suppress UCN losses in material containers.
[3] UCN trajectories are ballistic and considerably affected by gravity.
[4] Many review articles exist about Dark Matter. A concise review article is found, e.g., in the PDG's Review of Particle Properties [10]

## 2 UCN sources

Since 2011, PSI operates a high intensity source of UCN [11–17]. There was a steep learning curve in the operation of the very complex source system and the UCN output has been increasing over the years of operation. In 2015, the PSI UCN source became the world-leading UCN source in terms of number of UCN stored in a given experimental volume, see Fig. 1. For the purpose of source comparisons, a dedicated test-bottle was built [18] and sent to all relevant UCN sources to do comparative measurements [19]. This is very important because often only UCN densities were compared that were measured in different experimental configurations, e.g. with different size bottles of different materials. In an experiment, usually only the total number of stored and later counted UCN is important which is obviously depending on UCN density times a given volume. For instance, a specific source might be good to fill a small volume of 1liter to a relatively high density, however, by dilution, a large volume of 1 m$^3$ only to a small fraction of that density. The PSI UCN source is with its 2m$^3$ large intermediate storage vessel (see below, item 5 in Fig. 2) designed to also fill large volumes to high densities. Our 'standard UCN bottle' has a volume of 32l which was chosen to well represent the volume of typical planned neutron EDM experiments.

The lay-out of the UCN source at PSI is displayed in Fig. 2. The source is optimized to run in a macro-pulse scheme: it can take the full power proton beam from the ring cyclotron (590 MeV, 2.4 mA, 1.4 MW, not attenuated by the pion production targets M and E) for up

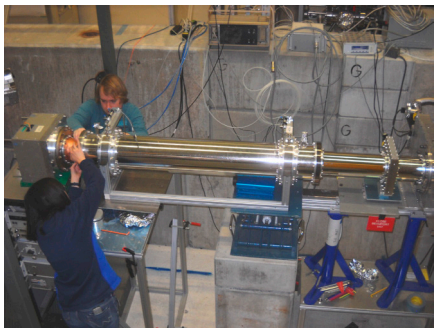
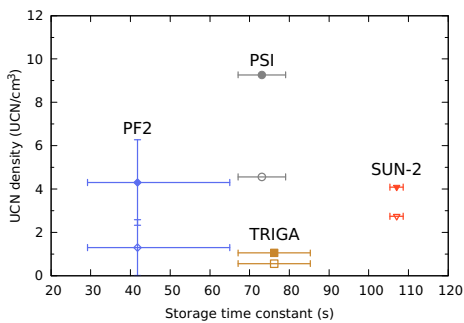

**Figure 1**

a) Picture of the 'standard UCN storage bottle' [18] in use for UCN density measurements in PSI UCN area West-1. b) UCN densities versus storage time at several UCN facilities in the world. UCN densities are given for storage times of 50 s (full symbols) and 100s (open symbols). Longer measured storage time constants correspond to lower average energies of the stored UCN spectra, thus lower wall collision rates and, accordingly, lower loss rates. PF-2 is the ILL facility based on the UCN turbine, SUN-2 is the new liquid-helium based installation at ILL, TRIGA is the channel-D UCN source at University of Mainz TRIGA reactor. (Figure reproduced with permission from [19].)

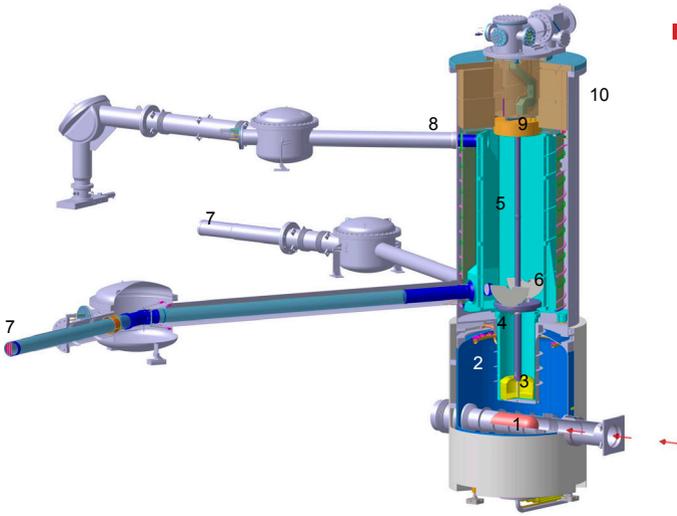

**Figure 2**
Cut-away drawing of the UCN source with indicated parts relevant to UCN production and transport (see text). 1 - spallation target, 2 - heavy water moderator tank, 3 - solid deuterium vessel, 4 - vertical guide, 5 - storage vessel, 6 - storage vessel flapper valve, 7 - UCN guide West-1 and South, 8 - upper UCN guide West-2, 9 - cryo pump, 10 - UCN source tank.

to 8 s with a duty factor slightly below 3%, limited by the average allowed beam current of up to 60μA. The proton beam hits a spallation target similar to the one of SINQ but horizontally mounted and with a larger diameter to distribute the higher peak beam power over the larger entrance window. The UCN target is cooled by heavy water and located inside a $D_2O$ thermal moderator.

The cold source in the center of the system consists of up to 30 liters of solid deuterium ($sD_2$) at 5 K to enhance the cold neutron flux and allow for down-scattering of cold neutrons to UCN. The latter is possible via phonon excitation while the reverse process is suppressed by the cold temperature of the solid. We thus have a super-thermal source of UCN [20]. The produced neutrons are transported with three UCN guides to two experimental areas, South and West. Figure 3 shows the UCN counts in a detector at the beamport West-1 as a function of time after a proton pulse.

Solid $D_2$ as UCN converter has first been studied in the 1980s [21,22] and received new attention later in the 1990s [23–25]. After initial demonstration experiments in Gatchina [26], a spallation-driven $sD_2$ UCN source prototype was first realized in Los Alamos [27]. Today, besides the UCN source at PSI, LANL and the TRIGA at Mainz university operate UCN sources based on $sD_2$. Other reactor sources are being prepared for the PULSTAR reactor at NCSU, USA, and for the FRM-2 in Munich. The work-horse for many UCN-physics experiments has for many years been the PF-2 instrument at the ILL Grenoble [28] based on mechanical deceleration of very cold neutrons. Other sources are in their test phases or under construction, e.g. those based on superfluid helium at ILL, TRIUMF, and PNPI. Cold and isotopically purified superfluid $^4$He presents an ideal superthermal source [20]. Its UCN production rate is about an order of magnitude smaller than that of solid $D_2$. However, UCN can live in sufficiently cold superfluid $^4$He essentially up to their β decay lifetime and can thus be accumulated over hundreds of seconds which is 3-4 orders of magnitude longer than in solid $D_2$. Such ideal conditions in helium are achieved at 0.5K and the losses grow rapidly with temperature proportional to T[7]. It is technically easier to cool $D_2$ to 5 K. Maintaining cold superfluid He next to an intense neutron source still needs to be demonstrated. Two routes are being pursued: One approach is to mount the source onto an external beamline, by that reducing the external heat input but sacrificing neutron flux. This is pursued at ILL [29]. The other one is to mount the UCN production volume close to a spallation source (at TRIUMF [30]) or to a reactor source (at PNPI [31]) in a very large neutron flux but compromise with the temperature, allowing for 1-1.2 K and reduced UCN lifetimes. At LANL and at PSI, there is now more than 10 years of experience in operating solid $D_2$ based, high flux neutron sources while the superfluid helium sources are still in exploratory stages.

## 3 Searching for a permanent electric dipole moment of the neutron

The highest impact experiment performed with UCN is the search for an electric dipole moment (EDM) of the neutron (nEDM). This endeavor started in 1950 with Smith, Purcell and Ramsey [32] and has been improved in sensitivity since then by more than 6 orders of magnitude [33]. While the spin−1/2 neutron has a well-known magnetic dipole moment (see section 4.1 below) which due to its interaction with magnetic fields makes it a superb probe of magnetic structure in neutron scattering, up to now no finite permanent electric dipole moment interacting with electric fields could be measured.

A finite EDM violates parity and time reversal symmetries [34–36], as illustrated in Fig. 4. The interaction energy of the electric dipole with the electric field is just the product of the two. The EDM must be either along or opposite to the direction of the spin vector - there is no other choice in a spin−1/2 system. Under parity (spatial inversion) the electric field changes sign while the spin is an axial vector.[5] Vice versa, under time reversal, the spin changes sign while the electric field is unaffected. This means that before and after the transformation, the contribution of this term changes sign. Thus, the system has a different energy and cannot be invariant under this transformation.[6]

In quantum field theory, time reversal T is intimately connected to CP violation via the CPT theorem [37] – with CP being the combination of charge conjugation and parity transformation and representing the symmetry between matter and anti-matter. Thus, a search for the nEDM is a search for an additional source of asymmetry between matter and anti-matter which is urgently needed in order to explain the apparent asymmetry of our universe – in which we have no primordial anti-matter at all, despite usually assuming a symmetric big-bang. The SM itself offers two sources of CP violation. The first one is a complex phase in the Cabibbo-Kobayashi-Maskawa (CKM) matrix which describes all the CP violating effects in the weak interaction. The strong interaction of the SM, described by quantum chromo-dynamics (QCD) also allows for a CP violating term in its Lagrangian, the so-called $\theta$-term. This is a named puzzle, 'the strong CP-problem', that this $\theta$-parameter appears to be unnaturally small, $\theta < 10^{-10}$. We know that it has to be so small because otherwise we would have already found a finite nEDM [38]. It could actually still be exactly zero. This is puzzling because the $\theta$ has a well-defined number of parameters and all allowed parameters are sizeable – but this one. There is no simple explanation within the SM why this parameter would be so small or even zero. The problem can be elegantly solved in theories beyond the SM, most prominently by the introduction of a symmetry and its breaking at the cost of a new particle, the so-called axion [39,40]. We will come back to axions which are at the same time top-ranking candidates for Dark Matter, see e.g. [41]. It is always attractive when theoretical solutions solve more than one problem at a time. However, no axion has been found yet and so its existence remains speculative and up to experiments to clarify.

---

[5] This is easiest seen for an electric field produced by a plate capacitor with the plates symmetric above and below the z = 0 plane. A spatial inversion will exchange the plates and with them the charges and thus reverse the electric field. The spin transforms like an angular momentum r × p where both r and p change sign under spatial inversion.

[6] The reader can immediately see that the interaction of the magnetic dipole moment with a magnetic field conserves parity and time reversal: The magnetic field is an axial vector as the magnetic moment, which is proportional to the spin vector. Therefore, both flip sign under T and both do not under P, and their product remains constant in both cases.

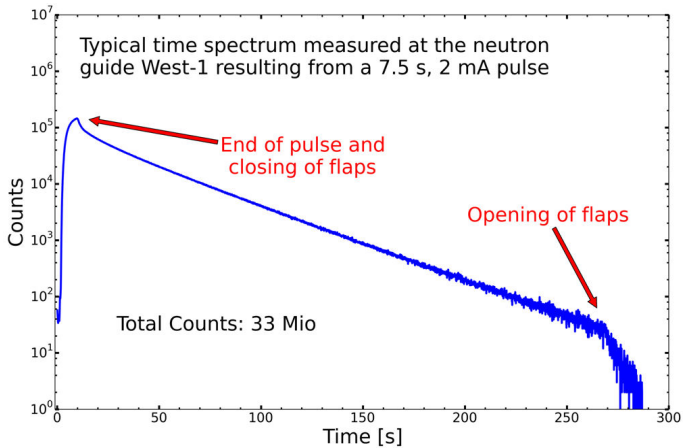

**Figure 3**
Time structure of UCN delivered to an experiment typically every 300 s. Closing and opening of the flaps refers to the central flapper valves, number 6 in Fig. 2. The flaps are open before the proton beam hits the spallation target and their closing time is optimized with respect to the end of the proton pulse to give the maximum number of UCN to the experiments. The displayed measured exponential decay of the UCN count rate has a time constant of about 30s reflecting the emptying time of the central storage vessel through the West-1 guide into the UCN detector. If the shutters to the UCN guides remain closed on the storage vessel, the storage time constant is about 90 s in the volume, number 5 of Fig. 2. At the end of a cycle, the flaps are re-opened to be ready for the next proton beam pulse.

---

[7] There are many excellent textbooks on the Standard Model and the CKM mechanism. A concise overview is found in the review section of the Review of Particle Physics [10].

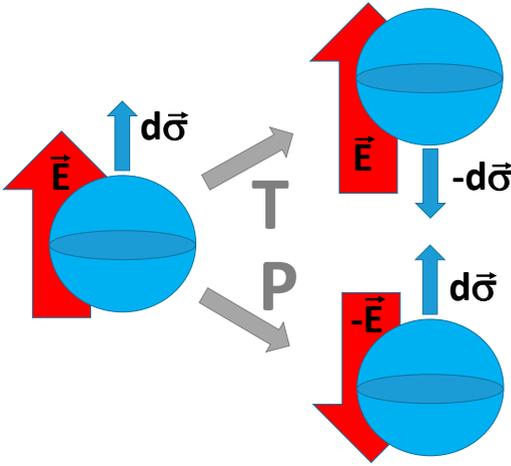

**Figure 4**
A finite electric dipole moment violates, both, parity (P) and time reversal (T) symmetries. The EDM can only be aligned or anti-aligned to the spin (σ) which is the only possible direction in the spin–1/2 neutron. The interaction energy of an EDM must be proportional to the product of the vectors of electric field (E) and EDM (dσ).

Figure 5 displays the continuous improvement of the nEDM limits, together with our latest result and our new experiment's sensitivity. Major drivers for progress with nEDM experiments always were increased statistical sensitivity due to better neutron sources and improvements of magnetic field control to suppress systematic effects due to the large interaction of the magnetic dipole moment with unavoidable magnetic fields. It didn't take long after the establishment of UCN that their application in nEDM experiments paid off. Especially important was the pioneering work with UCN at the Leningrad Nuclear Physics Institute (LNPI), e.g. [43], that in the late 1970s and in the 1980s established many of the techniques such as the first use of a double neutron precession chamber which finds its application again in our next n2EDM experiment at PSI, but also in other planned experiments of our competitors. Another very important step was the introduction of a cohabiting $^{199}$Hg magnetometer in the mid 1990s [44] by the Sussex-RAL-ILL collaboration.

Over time the available parameter space for viable theoretical models, giving rise to additional CP violation and predicting sizable nEDM, was continuously shrinking to remain compatible with the results of the measurements. While today still no finite EDM has been found, improved measurements of the nEDM remain top priority experiments in par-ticle physics. The search for the neutron EDM, in the course of its history, is said to be the one experimental approach which "excluded the largest number of theories and models"[8].

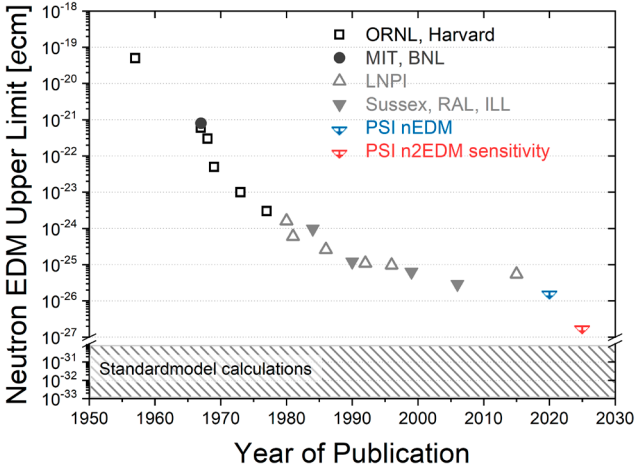

**Figure 5**
History of neutron EDM measurements as function of publication year, with the present best result from PSI [33] and the projection for n2EDM [42], if no signal would be observed. The transition from neutron beam experiments to UCN measurements was in 1980, starting with LNPI. The Standard Model range is estimated from the known CP violation in the weak interaction via quark mixing. It does not account for the possible CP violation in the SM's QCD sector. This contribution is actually limited by the limit of the neutron EDM itself. Therefore, the nEDM can also always be interpreted as a measurement of the θ-term of QCD which is the last unknown parameter of the SM which could still be exactly zero.

The international nEDM collaboration at PSI has taken high quality data 2015/16 and performed additional systematics measurements in 2017. Figure 6 shows a sketch of the nEDM experimental setup. Neutrons from the PSI UCN source are fully polarized after passing a 5 T magnetic field barrier, they are guided into a precession chamber inside a vacuum tank within an assembly of magnetic shields. In this chamber, well defined magnetic and

---

[8] This statement is attributed to N. F. Ramsey and J. M. Pendlebury. They are the individuals who have realized the largest progress on neutron EDM experiments to date. J. M. Pendlebury was a long-term member of the PSI particle physics committee and later joined the nEDM effort at PSI until he passed away in 2015. The 2015 update of the previous nEDM result was published by our collaboration in his honor [45].

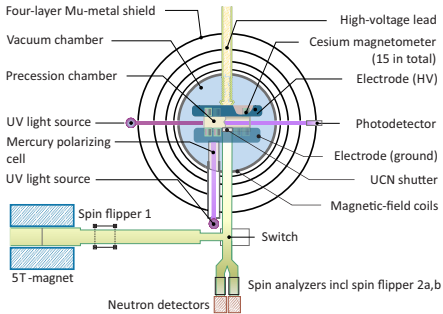 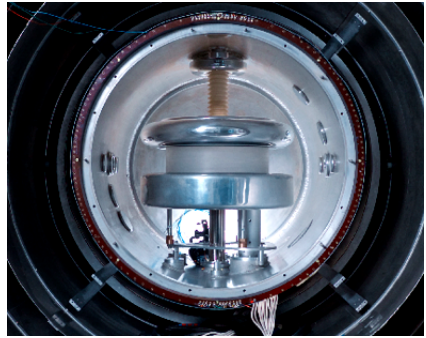

**Figure 6**

View of the nEDM apparatus at PSI with which the present best limit has been measured. The left figure (reproduced from [33]) shows all essential components of the apparatus. Its inner part can be seen on the photograph (on the right, by Markus Fischer, PSI). The diameter of the precession chamber is about 0.5m, the vacuum chamber has about 1m diameter and the outer diameter of the 4-layer Mu-metal shield is 2 m.

electric fields are applied to the stored neutrons. Their spins undergo a procedure of flips according to Ramsey's method of separated oscillatory fields. The procedure uses 3 minutes of free spin precession between two π/2 spin flips. This is quite a long free precession time for magnetic resonance experiments and key for the experimental sensitivity. The experiment measures the dependence of the phase accumulated during this precession on the applied electric field. This phase is encoded in the population of the spin up and spin down states after the second π/2 flip, after which the UCN are selected for their spin states and detected.

The data has been analyzed extremely carefully – over two years by two independent teams and with different blinding offsets for their respective data-sets – making sure not to bias the analysis with whatever the expectations for the result might be. Meanwhile the collaboration has unblinded the analysis and found perfect agreement of the independent analyses. The result of $d_n = (0.0 \pm 1.1_{stat} \pm 0.2_{syst}) \times 10^{-26}$ ecm gave no indication yet of a finite neutron EDM establishing a new world-best limit of $d_n < 1.8 \times 10^{-26}$ ecm (90% C.L.) [33]. The result was presented for the first time to the public in January 2020 at the users meeting of the particle physics community at PSI. The scientific impact is immediate, the parameter space of models of CP violation beyond the Standard Model was further squeezed.

From the experimental point most impressive is the success in reducing systematic uncertainties by more than a factor of five to less than $2 \times 10^{-27}$ ecm which gives good confidence for achieving the goals of the n2EDM project. n2EDM is the new folllow-up experiment by the same international collaboration at PSI, aiming at an order of magnitude improved sensitivity to the neutron EDM as the next goal. Construction of the new apparatus is ongoing in area South of the UCN facility. Figure 7 shows the completed setup of the magnetically shielded room (MSR) of n2EDM before a wooden house was built around it as thermal enclosure. Temperature stability is one key aspect to guarantee proper functioning of the MSR. Stabilizing the temperature around the MSR to ±1 K leads to millikelvin stability inside of the room over many hours. Another key is the reduction of any ambient magnetic field drifts around the MSR as much as possible. This is taken care of by an active magnetic stabilization (AMS) system using multiple coils and many fluxgate sensors for magnetic field measurements and feedback stabilization of the field. An innovative coil design method was developed [46] and implemented with a small scale prototype of the AMS at ETH Zurich. The commissioning of the full n2EDM setup is planned for 2021. Data taking will afterwards take another couple of years before high-quality data for a new result will be available.

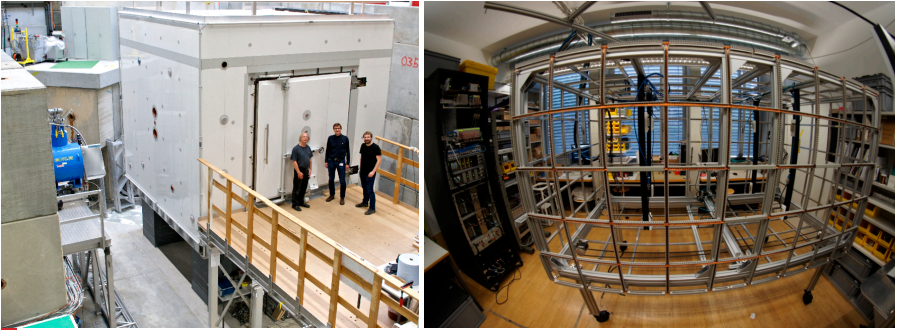

**Figure 7**
(Left) Picture (by Markus Fischer, PSI) of the authors together with the n2EDM technical coordinator Georg Bison in front of the just assembled magnetically shielded room (MSR) of n2EDM in November 2019. The MSR has 6 Mu-metal layers and an almost cubic inner room of 3 m side length. Meanwhile the MSR has been surrounded by a thermal enclosure to stabilize the temperature for the experiment. The system will provide the best magnetically controlled large volume in Switzerland and perhaps worldwide. The blue magnet on the left is the same 5 T polarizer magnet as seen in Fig. 6 on the left. (Right) Scaled down prototype of the active magnetic shield (AMS) at the ETH Zurich (photo: Michal Rawlik, ETHZ). The AMS for n2EDM will be mounted in summer 2020, directly on the inside of the thermal enclosure around the MSR.

While the PSI UCN area South is dedicated to the n2EDM effort, area West of the UCN source provides two more beamlines with increasing demand for cross-section measurements, detector tests and developments, UCN depolarization studies, UCN source studies and several more. The West-1 beamline is also a good candidate to host future experiments concerning neutron lifetime, neutron decay or neutron oscillations to invisible states. For the last topic, a letter of intent for an experiment was just presented at the users' meeting in January 2020. Invisible states here refers to particles that would also be well-motivated candidates for Dark Matter. A tiny, hypothetical interaction could allow neutrons to oscillate to mass-degenerate particles, so-called mirror-neutrons. Those would have almost no interaction with ordinary matter and disappear from the apparatus.

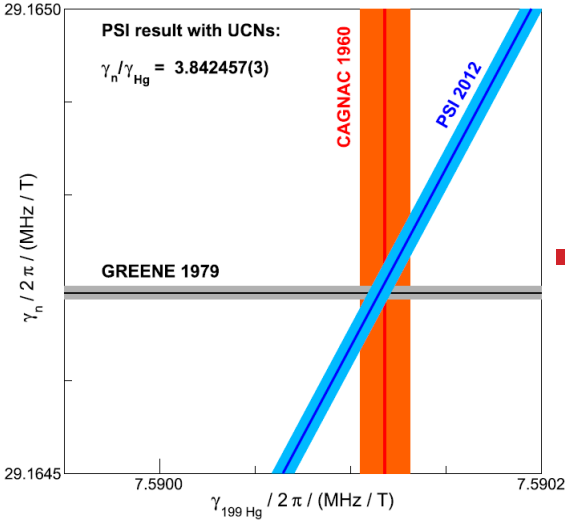

**Figure 8**
Precision comparison of magnetic moments of the neutron and 199Hg, reproduced from [49].

## 4 Other results from the nEDM collaboration

### 4.1 The magnetic moment of the neutron

The experimental art of the search for the neutron EDM is to a large extent in the control of the magnetic fields with which the magnetic moment of the neutron interacts. As these magnetic fields can never be perfectly homogenous or perfectly stable in time they must be measured online with high precision and accuracy within the experiment itself. The nEDM experiment exploited different magnetometry systems, among them laser-optically pumped Cs magnetometers [47] surrounding the UCN spin precession chamber and a cohabiting $^{199}$Hg magnetometer occupying the same volume during the measurement as the UCN, see Fig. 6. By means of these magnetometers and a careful offlinemapping procedure, the magnetic fields in the experiment have been determined with unprecedented precision [48]. As a side effect, it becomes possible to link the magnetic moments of the neutron and the $^{199}$Hg atoms by measuring the ratio of their Larmor precession frequencies and controlling all systematic effects that can shift the frequency ratio. Figure 8 displays the result [49] of the precision measurement of the ratio of gyromagnetic ratios of the two species. As can be seen, our result is compatible with the still more precise measurement of the neutron magnetic moment and the less precise measurement of the one of $^{199}$Hg. In the future, this ratio will be further improved by the n2EDM experiment. In a separate measurement, one can link the $^{199}$Hg magnetic moment to the one of $^{3}$He and by that reference the neutron magnetic moment to the one of $^{3}$He and thereby also to the proton. Further improvement of the precision on the neutron magnetic moment is thus possible and can be delivered as a byproduct of the experimental n2EDM program. While measurements of the neutron magnetic moment with UCN storage bottles will likely also in the future not match the presently thousand times better precision of the highly sophisticated penning-trap experiments for the proton [50] (nor for 3He), it is a fundamental constant and important benchmark for experiments and, possibly, for future precision QCD lattice calculations of baryon properties.

## 4.2 Spin echo spectroscopy with UCN

For UCN stored in a container, the energy spectrum is limited by the container's wall potential. As the energy for individual UCN is conserved and spectral groups do not mix during storage, spectra almost only evolve because of UCN energy dependent loss mechanisms. This means in particular, that the average height of UCN in a containment is determined by their energy and, vice versa, that the average height encodes the UCN energy. Based on these ideas, the nEDM collaboration at PSI has pioneered a spin-echo technique with UCN that allows to reconstruct the energy spectrum of stored UCN [51].

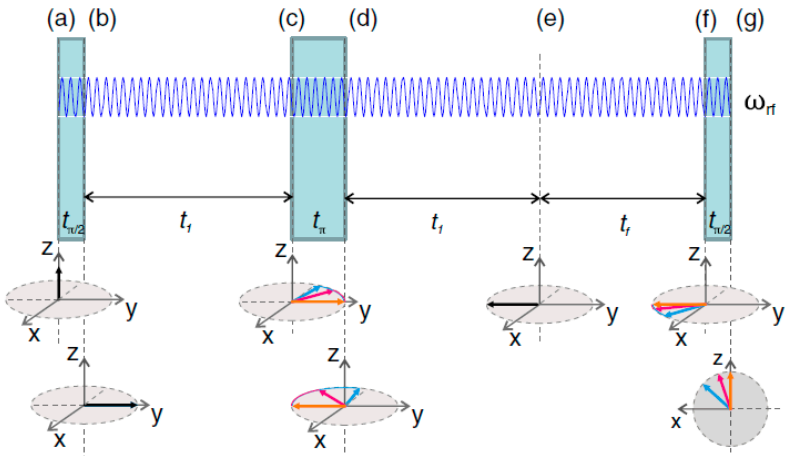

**Figure 9**
Scheme of the neutron spin-echo sequence with subsequent π/2, π and π/2 pulses and spin projections for three different UCN energies in a magnetic field with a vertical gradient, see text. Figure reproduced with permission from [51].

The magnetometry of the experiment gives access to the control of magnetic field gradients in the apparatus. Applying, for example, a linear, vertical magnetic field gradient allows to correlate the average Larmor precession frequency with the average height of the UCN and thus their energy.

Figure 9 displays a spin-echo sequence for stored UCN. In addition to the two π/2 pulses of a Ramsey measurement, a π pulse is introduced inbetween. After the first π/2 pulse, the initial UCN polarization along z is oriented along the y axis. During the free spin precession, the magnetic field gradient leads to a dephasing of the spins, as different UCN energy groups (three are displayed) sample the magnetic field on average at different height and therefore differently. The spins of the UCN sampling on average higher fields run ahead of those sampling lower fields. After a time $t_1$ a π pulse rotation around the x axis changes the sequence and the faster precessing spins are now behind the slower ones. After another time $t_1$, the faster ones catched up and the spins are in phase again. Applying the second π/2 pulse at this time would result in maximal UCN polarization (that is analyzed along z), not affected by the vertical magnetic field gradient. Adding a variable time $t_f$ before the second π/2 pulse leads to new dephasing. Variation of $t_f$ then allows to scan the UCN dephasing and thus to determine with an unfolding procedure the number of UCN per energy bin.

Obviously, the method allows to extract spectral information for known magnetic field gradients. In turn, it also allows obtaining vertical magnetic field information for known UCN spectra. Moreover, one can disentangle different contributions that lead to UCN spin polarization loss during storage as field dephasing effects can be distinguished from irreversible depolarization in wall collisions.

### 4.3 Searches for 'exotic physics'

The term 'exotic physics' is a particle physics slang referring to physics beyond the SM which at the same time is not main-stream. To give a first example, the SM is a quantum field theory (QFT) and so are most well-motivated extensions of it. Lorentz Invariance (LI) is at the very foundation of QFT and its violation (LIV) not easily incorporated in a theory. Nevertheless, especially those pillars of our best theories should be experimentally tested with the highest possible precision. On the one hand, any observation of LIV would break a pillar of QFT and by that shake the foundations of modern physics. On the other hand, LI symmetry assumes isotropy of space which is assumed to be given at large but is certainly not given with gravity in our laboratories, in our solar system or our galaxy. Theorists are building models or effective theories that can include gravitational effects and can allow for LIV, see e.g. [52]. One important feature of that approach is that experiments can now test and constrain the parameters of such models, and very different experiments can constrain the same parameters and be systematically compared.

The measurement of the ratio of gyromagnetic ratios of neutron and $^{199}$Hg, as used to determine the magnetic moments, see section 4.1, is a sensitive clock comparison. Moreover, in the nEDM setup, the vertical axis in the laboratory is the magnetic field axis around which the Larmor precession takes place. As Earth rotates around itself once per siderial day and around the sun over the year, this axis, fixed in the laboratory, samples different directions in space. One simple setting for a LIV effect would be the existence of a preferred direction in space (thus violating isotropy), e.g. by a vector field that could act on the precessing spins, almost like a magnetic field. The experimental signature would be a diurnal or annual modulation of the ratio of precession frequencies as neutron and $^{199}$Hg spins would be affected differently. This effect has been searched for but was not found. Therefore limits were set by the nEDM collaboration on early data, still measured at the ILL, on the magnetic effect [53] as well as on a slightly more complicated EDM signal effect [54] constraining best the relevant parameters of the LIV parametrization. Updates based on the precision data set from PSI are being worked on.

LIV is not the only possible source of oscillating EDM signals. The already mentioned axions, or in more general terms 'axion-like particles' (ALPs), could be (but don't have to be) ultra-light and still be viable candidates for Dark Matter, see [41]. Being ultra-light and at the same time making up for the amount of Dark Matter in our Universe would result in very large quantities of such particles. The tiny mass of ALPs comes together with very long Compton wavelengths and very low Compton frequencies, and the abundant particles could be just described as a classical, coherently oscillating field.[9] ALPs could induce EDMs of neutrons via their coupling to gluons (as with the QCD θ-term) and those EDMs would be oscillating at the Compton frequency of the axion field. As one cannot a-priori know this frequency, one has to perform a broad search at all accessible frequencies. For the nEDM experiment, this is the range of inverse time scales of the measurements from single cylce duration of a few minutes to several years of data taking, covering some six orders of magnitude. We have performed a search for such oscillating signals in two data sets, an older one from the ILL and our newer one from PSI. Because the PSI data at the time of this ALPs analysis was still blinded, it could for technical reasons in the analysis not yet be used for long times. Because, in turn, the absolute time stamping of the ILL data was not sufficiently good, that data set could not be used for short times. Figure 10 shows the results for the two time bases which nicely complement each other. No signal above the expected noise could be found and thus signals as a function of frequency have been excluded down to certain weak coupling strengths [56]. The axion frequency is shown on the upper horizontal axis and equivalently the axion mass on the lower

---

[9] The fact that the oscillation is coherent is not trivial, see e.g. [55]. It follows from cosmological arguments that the axion field started at an almost constant value everywhere in the observable universe after inflation, because inflation exponentially suppressed all spatial gradients of the field. Starting everywhere from the same value, it behaves like a coherently oscillating classical scalar field. The fact that the gradients must be small, translates into a small velocity, making the axions a viable Cold Dark Matter candidate. Of course, the coherence is not perfect and the coherence time is limited to some $10^6$ periods.

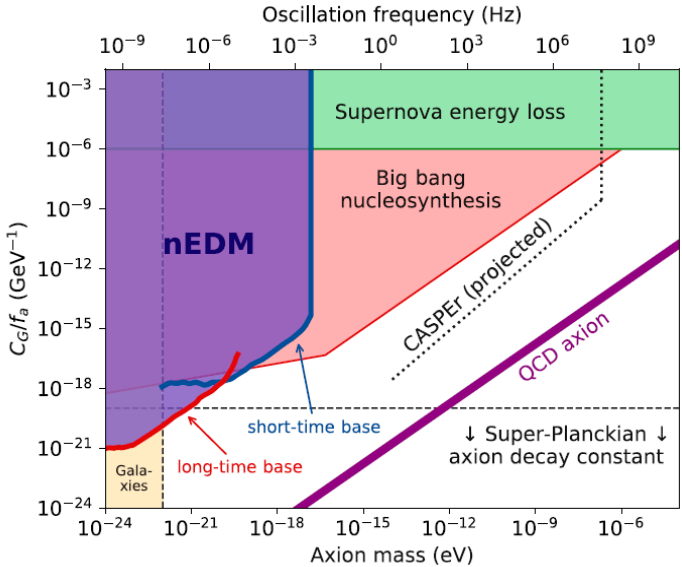

**Figure 10**

First laboratory exclusion of axion-like particle Dark Matter that would generate oscillating neutron EDM signals. The normalized coupling strength of ALPs to gluons in the neutron is shown as a function of ALP mass or frequency, respectively. In addition to the nEDM exclusion region, exclusion regions from astrophysical and cosmological observations and arguments are shown. A band is shown where the QCD axion would be located, illustrating that the nEDM is so far sensitive to ALPs but not to the effect of QCD axions. CASPEr is another magnetometry project that aims at the indicated sensitivity. Figure reproduced from [56], see there and text for more details.

horizontal axis. The vertical axis shows the coupling strength to the gluons in the neutrons normalized to the so-called axion decay constant $f_A$. The nEDM result is the first laboratory limit on the mass and coupling of such particles. One of the astrophysical limits labeled 'Galaxies' is from the fact that ALPs as light as $10^{-22}$ eV have a wavelength of the size of dwarf galaxies. As Dark Matter also exists in those galaxies, the wavelength of ALPs that make up for all Dark Matter cannot become larger. It is exceptional that a laboratory experiment can exclude such regions and even beyond. Equally amazing is the fact that the experiment probes, in case of the ultralight axion masses, axion decay constants larger than the Planck mass of almost $10^{19}$ GeV. Thus, this is an experiment using neV neutrons, searching for ultra-light $10^{-22}$ eV axion-like particles while being sensitive to extremely high-energy, Planck mass-scale physics. The former ETH-PSI doctoral student Michal Rawlik was awarded the CHIPP prize 2019 for the best PhD student in Swiss particle physics, to a large extent for his work on this analysis, but also for his innovative technique [46] to design the external magnetic field compensation coil system for the n2EDM experiment.

Axion-like particles are not restricted to being ultra-light. They could also, for example, be in a range of meV to eV, corresponding to Compton wavelengths of mm to µm. They could not only induce a neutron EDM, they could also mediate additional spin-dependent interactions, e.g. between polarized neutrons and unpolarized nuclei. This hypothetical interaction would be short-range because of the massive mediator, with a characteristic interaction length given by the Compton wavelenght of the ALPs. It can be parametrized with a Yukawa potential, with a coupling strength and the interaction length as parameters.

The nEDM experiment provides an ideal configuration to search for such hypothetical interactions. Polarized UCN precess in the storage chamber of the setup (see Fig. 6) and interact with the unpolarized nuclei of the walls. UCN very close to the lower or to the upper electrode would be most influenced but would display opposite spin phase shifts. If the UCN distribution would be homogeneous, the effects of the lower and upper electrode would, thus, exactly cancel. However, the UCN distribution is not homogenous. In fact, the UCN center of mass is 3-4 mm below the central plane of the 120 mm high chamber. This is because the slow UCN are significantly affected by gravity. By contrast, the $^{199}$Hg atoms are thermal and fill the chamber homogeneously. Therefore, they are inert against any such spin dependent effect described for the neutrons. As the UCN sample the region close to the bottom electrode more often, this effect dominates. If the main vertical magnetic field of the experiment is reversed and the Larmor precession changed to opposite direction, any effect that was adding to the phase of the UCN spins would now be subtracted. In such a way, a potential signal could be isolated as the system's magnetometry guarantees magnetic field control. Such a search has been performed by the nEDM collaboration on early PSI data. No signal was found and the best limit at that time was derived [57], an exclusion plot in a plane of coupling strength versus interaction range. The result has been superseded in the meantime by other experiments using polarized atoms rather than neutrons, however, the nEDM collaboration is analyzing its new data for another order of magnitude improvement, aiming again for the lead.

One might wonder why searching and often finding zero is worth the effort: 'Why measuring zero to higher precision ever and ever again?', the eminent neutron physicist Geoffrey Greene recently asked. He continued to give the answer himself: 'Because we can'. The attempt to understand Nature at the most fundamental level is driving this research. Many limitations of our Standard Model as discussed already in section 1 are known, together with a large variety of possible extensions – known and yet unknown – of the SM. Among them, reality itself is hiding. The 'Because we can' is perhaps even a 'Because we must'. We are determined to decypher Nature's most basic laws and each time we find zero at better precision is like a little discovery – of what Nature is not, although it could have been. At some point, deviations of the SM will show up and open an era of new physics. We need to push the boundaries with experiments using cutting edge technology and theoretical guidance. We do not simply measure everything as precisely as possible but try to concentrate on the most promising systems. Such experiments can give access to very high energy scales, complementary to direct searches for new particles at high energy colliders, such as the LHC at CERN. In some cases, the experiments reach far beyond the energy of existing and even future colliders. Neutrons, and especially UCN, have a lot to offer in this respect.

## 5 Other UCN activities worldwide

The research with ultracold neutrons at PSI is not an isolated activity but part of a growing world-wide research community. Activities concerning the development, construction and operation of UCN sources have been mentioned in section 2. Several of the new source projects aim at measuring the neutron electric dipole moment, in direct competition with our effort at PSI. Mainly the sources at ILL (PF-2) and at Los Alamos have also delivered physics results on different topics. Both have made major contributions to the measurement of the β-decay lifetime of neutrons[10] with UCN, the value of which is presently puzzling many physicists because of a discrepancy in the results obtained from experiments using stored UCN and, respectively, a cold neutron beam [59]. While at PF-2, both, experiments with material bottles have been brought to perfection [60] and with magnetic storage have been pioneered, the UCNτ experiment at LANL is currently leading the field with a large magnetic storage bottle [61]. Because neutron β-decay has two parameters in the Standard Model that need to be determined, namely the ratio of axial-vector to vector coupling $g_A/g_V$ [11] and the matrix element of the CKM matrix describing the mixing of u and d quarks, $V_{ud}$, one usually combines the result of the lifetime measurements with decay-correlations, for instance the correlation A of the neutron polarization and the momentum of the decay electron in the decay $n \dashrightarrow p\bar{\nu}_e e^-$. This allows for sensitive tests of the Standard Model when, e.g. combined with other elements of the CKM matrix, in particular $V_{us}$, to test its unitarity. At LANL, the decay correlation A has been measured with the UCNA experiment [62] providing very different systematics compared to the latest measurement on a cold neutron beam at ILL, which has still considerably higher precision [63].

Another activity with UCN that has gained considerable attention is the spectroscopy of UCN gravitational quantum bound states above a mirror, see e.g. [64]. Besides being a wonderful, text-book model-system, in principle, the precision spectroscopy of these states is a sensitive tool to search for additional short range interactions between the neutron and the material of the mirror. While experiments with cold atoms reach higher precision, they might have larger systematic issues, as the atoms experience considerably larger interactions of electromagnetic origin on a surface that need to be corrected for.

---

[10] The precise value of the neutron lifetime is a key parameter in big-bang nucleosynthesis (BBN). The primordial helium abundance is directly related to the neutron lifetime which determines the number of free neutrons still available at the time of formation of the first nuclei, see e.g. [58].

[11] The weak interaction coupling in the SM is purely V–A, vector minus axial-vector. The terminology refers to the Lorentz structure of the interaction. See standard particle physics textbooks for reference.

With a continued improvement of UCN sources and new technological developments, from UCN optics via magnetometry and magnetic field stabilization to novel UCN detection schemes, the field will continue to push the boundaries of what is known and possible.

## 6 Acknowledgements

The authors acknowledge valuable contributions of the present and previous members of the UCN physics group at PSI, of the UCN source operation group, and of the international nEDM collaboration at PSI. The experiments of the nEDM collaboration have been supported by many national funding agencies across Europe and by the ERC. The authors are especially grateful for the continued support of PSI, ETH Zurich and the Swiss National Science Foundation. We also gratefully acknowledge the support of many service groups at PSI and the outstanding efforts to provide the HIPA facility with an unmatched proton beam performance on which everything else builds up.